\documentclass[]{spie}  
 
\usepackage{amsmath,amsfonts,amssymb}
\usepackage{graphicx}
\usepackage{siunitx}
\usepackage[colorlinks=true, allcolors=blue]{hyperref}
\usepackage{lipsum} 
\usepackage{multirow}
\usepackage{caption} 
\newcommand\Tstrut{\rule{0pt}{2.6ex}}         

\title{Studies of Systematic Uncertainties for Simons Observatory: Polarization Modulator Related Effects}

\author[a]{Maria Salatino}
\author[b]{Jacob Lashner}
\author[c]{Martina Gerbino}
\author[d]{Sara M. Simon}
\author[b]{Joy Didier}
\author[e]{Aamir Ali}

\author[e,f]{Peter C. Ashton}
\author[g]{Sean Bryan}
\author[e]{Yuji Chinone}
\author[d]{Kevin Coughlin}
\author[h]{Kevin T. Crowley}
\author[i]{Giulio Fabbian}
\author[l]{Nicholas Galitzki}
\author[e]{Neil Goeckner-Wald}
\author[d]{Joseph E. Golec}
\author[c]{Jon E. Gudmundsson}
\author[e,f]{Charles A. Hill}
\author[l]{Brian Keating}
\author[f,m]{Akito Kusaka} 
\author[e,f,n]{Adrian T. Lee}
\author[d]{Jeffrey McMahon}
\author[b]{Amber D. Miller}
\author[o]{Giuseppe Puglisi}
\author[p]{Christian L. Reichardt}
\author[l]{Grant Teply}
\author[q]{Zhilei Xu}
\author[q]{Ningfeng Zhu}

\affil[a]{AstroParticle and Cosmology Laboratory (APC), Paris Diderot University, 75013, Paris, France}
\affil[b]{Department of Physics and Astronomy, University of Southern California, Los Angeles, CA 90089, USA
}
\affil[c]{The Oskar Klein Centre for Cosmoparticle Physics, Department of Physics, Stockholm University, SE-106 91 Stockholm, Sweden}
\affil[d]{Department of Physics, University of Michigan, Ann Arbor, MI, USA}
\affil[e]{Department of Physics, University of California, Berkeley, Berkeley, CA, USA}
\affil[f]{Physics Division, Lawrence Berkeley National Laboratory, Berkeley, CA, USA}
\affil[g]{School of Electrical, Computer and Energy Engineering, Arizona State University, Tempe, AZ, USA}
\affil[h]{Department of Physics, Princeton University, Princeton, NJ, USA}
\affil[i]{Institut d'Astrophysique Spatiale, CNRS (UMR 8617), Univ. Paris-Sud, Universit\'{e} Paris-Saclay, B\^{a}t.121, 91405 Orsay, France}
\affil[l]{Department of Physics, University of California San Diego, La Jolla, CA, USA}
\affil[m]{Department of Physics, The University of Tokyo, Tokyo, Japan}
\affil[n]{Radio Astronomy Laboratory, University of California, Berkeley, Berkeley, CA 92093, USA}
\affil[o]{Department of Physics, Stanford University, Stanford, California, CA 94305, USA}
\affil[p]{School of Physics, University of Melbourne, Melbourne, Australia}
\affil[q]{Department of Physics \& Astronomy, University of Pennsylvania, Philadelphia, PA 19104, USA}

\authorinfo{Send correspondence to Maria Salatino. E-mail: maria.jokeating@gmail.com.}

\begin{document} 
\maketitle
\begin{abstract}
The Simons Observatory (SO) will observe the temperature and polarization anisotropies of the cosmic microwave background (CMB) over a wide range of frequencies (27 to 270~GHz) and angular scales by using both small ($\sim$0.5~m) and large ($\sim$6~m) aperture telescopes. The SO small aperture telescopes will target degree angular scales where the primordial B-mode polarization signal is expected to peak. The incoming polarization signal of the small aperture telescopes will be modulated by a cryogenic, continuously-rotating half-wave plate (CRHWP) to mitigate systematic effects arising from slowly varying noise and detector pair-differencing. In this paper, we present an assessment of some systematic effects arising from using a CRHWP in the SO small aperture systems. We focus on systematic effects associated with structural properties of the HWP and effects arising when operating a HWP, including the amplitude of the HWP synchronous signal (HWPSS),  and $I\rightarrow P$ (intensity to polarization) leakage that arises from detector non-linearity in the presence of a large HWPSS. We demonstrate our ability to simulate the impact of the aforementioned systematic effects in the time domain. This important step will inform mitigation strategies and design decisions to ensure that SO will meet its science goals.


\end{abstract}



\section{Introduction}

\label{sec:intro}  

The cosmic microwave background (CMB) is a rich source of cosmological information. Accurate measurements of the small angular scales (large multipole moments, or high $\ell$) would enable tightening constraints on the sum of neutrino masses, the number of relativistic species, dark energy equation of state, dark matter properties, and astrophysics of galaxy clusters. On large angular scales (low $\ell$), the detection of the faint primordial B-mode polarization signal could hold information about the energy scale of inflation, which is parameterized by the tensor-to-scalar ratio $r$. 
For ground-based observatories, there are two primary sources of contamination to the CMB polarization: polarized foreground contamination from synchrotron and dust emission and effects induced by atmospheric fluctuations. The latter becomes particularly important on large angular scales where the primordial B-mode signal from inflation is expected to peak.

Simons Observatory (SO) is a ground-based experiment that will be located in the Atacama desert in Chile. SO will observe the CMB sky in temperature and polarization across six bands with band centers ranging from 27~GHz to 270~GHz. It will cover a wide range of angular scales through the use of a 6~m diameter crossed-Dragone large aperture telescope (LAT) and several refractive $\sim$0.5~m small aperture telescopes (SATs). The wide coverage in frequency will be achieved through the use of multichroic polarization-sensitive detectors and will allow for the characterization and removal of the galactic foregrounds. The SATs will observe on degree scales and focus on measuring the polarized signal from inflation. Noise from long-time-scale atmospheric fluctuations --- so-called $1/f$ noise --- can overwhelm the polarized inflationary signal. To mitigate this effect, the SO SATs will employ cryogenic, continuously-rotating half-wave plates (CRHWPs) that modulate the polarized signal and isolate it from the largely unpolarized atmospheric fluctuations~\cite{Kusaka14,Essinger16,Takakura:2017ddx,EBEX17}\footnote{We note that the atmosphere is slightly circularly polarized~\cite{Hanany03, Spinelli11}, and a significantly lower amplitude of linear polarization is predicted. However, this is still below the current detection sensitivity.}. The LAT will not use any HWPs. Because CRHWPs are most effective at recovering polarization signals at large angular scales, the scientific return of featuring a HWP in the LAT for high-$\ell$
science is negligible compared to the complex design requirements for having a HWP. 

HWPs for millimeter-wavelength applications have been fabricated from both sapphire~\cite{Kusaka14,Hill:2016jhd,EBEX17} and silicon meta-material~\cite{Henderson16}. The main advantage of a silicon meta-material HWP is its high birefringence. The difference between the indices of refraction along the two propagation directions through the material are $\Delta n\simeq 1$ for meta-material compared to $\Delta n\simeq 0.3$ for sapphire. Additionally, silicon has a high index of refraction, which allows for a thinner HWP and thus less thermal emission. However, this advantage is less significant for cryogenic HWPs where the thermal emission is small, so, at cryogenic temperatures, the ease of manufacturing sapphire HWPs becomes more advantageous than the reduced loading offered by silicon meta-material HWPs. Thus, the cryogenic CRHWPs employed in SO will be manufactured from sapphire.

This paper will present systematic studies of the SO SAT CRHWPs, focusing on the case of a broadband-HWP
design for the mid-frequency (MF) 90/150 GHz channels. We explore the physical sources of these systematic effects, illustrate our ability to model these effects with a time-domain systematics pipeline (\texttt{s4cmb}\footnote{The code is publicly available at the following github link: \url{https://github.com/JulienPeloton/s4cmb}.}), and discuss how these studies have influenced the experiment design. In Sec.~\ref{sec:Mueller}, we derive the mathematical description of the optical properties of HWPs, including their dependence on frequency and the incident angle of the incoming light using the Mueller matrix formalism. SO considered both sapphire and silicon meta-material HWPs. A functional understanding of the systematic effects induced by each was valuable in selecting the HWP material, so we present results for both sapphire (Sec.~\ref{sec:sapphire_mueller}) and meta-material (Appendix~\ref{sec:metamaterial}) HWPs in this work. Next we use the sapphire SO HWP Mueller matrices to study differential properties. In Sec.~\ref{HWPSS_Intro}, we take the optical model and include its properties in time-ordered data (TOD) with an emphasis on systematic effects related to the HWPSS. In this calculation, we assume the baseline optical configuration for the SO SAT. In Sec.~\ref{sec:NL}, we investigate the resulting temperature to polarization ($I \rightarrow P$) leakage and nonlinear behavior in the detectors from the HWPSS.

This paper is part of a series of papers on the systematic and calibration studies for SO~\cite{crowley18,gallardo18,bryan18}. We are combining the detailed results of the full SO systematics and calibration studies into a comprehensive study that will be released in a series of future papers to the community for use in developing future CMB experiments such as CMB-S4.\cite{cmbs4}. In this paper, we take an in-depth look at a small number of HWP systematics and note that further information on HWP systematic effects that are not included in this paper will be included in the full systematics study papers.

\section{Mueller Matrix Model of a HWP}\label{sec:Mueller}
The Mueller matrix of a HWP of any material (sapphire, silicon meta-material or metal mesh \cite{Pisano08}) can be written as:
\begin{equation}\label{eq:Mueller}
M^{\mathrm{HWP},k}=\left(
  \begin{array}{cccc}
    t_k & \rho_k & 0 & 0 \\
    \rho_k & t_k  & 0 & 0 \\
    0 & 0 & c_k & -s_k \\
    0 & 0 & s_k & c_k \\
  \end{array}
\right)
\end{equation}
where $k=T$ or $R$ for the transmission and reflection, respectively~\cite{Moncelsi12,Zhang11}. Here the $\rho_k$ terms are proportional to the HWP differential transmission. The sign of $\rho_k$ determines whether the even or odd harmonic peaks in the HWP signal have a larger amplitude as it turns through a full rotation,
while its magnitude determines the difference between the amplitudes of the odd and even peaks. The $c_k$ and $s_k$ terms determine the modulation of the polarized signal. An ideal HWP has no reflection, so $t_k=1$, $c_k=-1$, and $\rho_k=s_k=0$. A non-ideal HWP is characterized by $c_k\neq-1$ and nonzero $\rho_k$ and $s_k$ terms\footnote{The Mueller matrix of an achromatic sapphire HWP has more nonzero terms, see Sec.~\ref{sec:HWP_vs_angle_and_nu}.}. The structure of the Mueller matrix is independent of the direction of light along the HWP's normal axis and thus has the same structure for both transmission and reflection. The exact value of the Mueller matrix components changes according to the Fresnel coefficients for reflection and transmission.

Depending on the HWP material, different methods can be employed to calculate the values of the Mueller matrix elements in Eq.~\ref{eq:Mueller}. For sapphire HWPs without anti-reflective (AR) coatings, the Mueller matrix components can be derived analytically by explicitly taking into account multiple reflections inside the material~\cite{Salatino17}, while HWPs both with and without AR coatings can be modeled via the transfer matrix model~\cite{Essinger13}. In the case of more complex material layers, such as silicon meta-material and metal mesh HWPs, electromagnetic simulations in High Frequency Structure Simulator (HFSS)~\cite{HFSS} or similar softwares are needed. HFSS enables considering scattering and diffraction effects in more complicated geometries where analytical models are not possible. 

When traversing a HWP, polarized light is rotated by twice the angle between the HWP optical axis and the polarization direction of the incoming light. Thus, if the HWP is rotating with frequency $f$, the polarized signal transmitted through the HWP is rotated at 2$f$. These signals are measured by polarization sensitive detectors at twice the rotation frequency, so the polarized signal is modulated at 4$f$. The differential properties of the HWP can polarize a fraction of the unpolarized signal. At normal and slant incidence, this signal rotates at the same frequency as the HWP and couples to the detectors at $2f$. At slant incidence, an additional component at $4f$ can be produced
by an inherent dependence on the HWP rotation angle in its Mueller matrix.


\subsection{Sapphire HWPs} \label{sec:sapphire_mueller}
	An actual HWP has imperfect behavior and an imperfect AR coating. We make use of the generalized transfer matrix method (TMM) to model the optical action of sapphire HWPs. The TMM determines the relation between the transverse-magnetic (TM) and transverse-electric (TE) components of the incoming and outgoing radiation at the two boundaries of an optical element (or a stack of elements) using the boundary conditions. The TMM has several advantages. This method deals directly with the amplitudes of the fields at the interfaces of the two layers, and thus automatically takes multiple reflections between layers into account. The TMM is easy to generalize to multi-layer elements like broadband achromatic HWPs (AHWPs) with AR coatings. Finally, the TMM can be easily applied to the more general case of arbitrary incidence angle and HWP rotation. Once the transfer matrix for the stack is computed, it is also possible to compute the corresponding Jones matrix and Mueller matrix\footnote{A public version of the code can be found at \url{https://github.com/tomessingerhileman/birefringent_transfer_matrix}, thanks to Tom Essinger-Hileman.} (see e.g.\cite{Abel50,Born99,Hecht87,Berreman72,Yeh79,Essinger13}).

For the SO HWP, we consider a three-layer AHWP built up from three individual HWPs with a two-layer AR coating (on each side) with thicknesses and indices of refraction matching the \textsc{Polarbear}-2 AHWP design, which are listed in Tab.~\ref{tab:AHWP}~\cite{Hill:2016jhd}. We note that the \textsc{Polarbear}-2 design is used here as a reference design of a realistic HWP to demonstrate the performance of our analysis framework and to estimate an approximate systematic level of a HWP in our system. The SO HWP may vary in design from the \textsc{Polarbear}-2 HWP design and will be cryogenic, meaning that the exact values of the indices of refraction and of the loss tangent of sapphire at cryogenic temperatures will differ. As the SO HWP design is refined, we will use this framework to assess its systematic performance as we do here for the \textsc{Polarbear}-2 HWP design. Possible losses inside the materials are modeled via a non-vanishing loss tangent. This is included as the imaginary part of the complex index of refraction $\tilde{n}=n\sqrt{1-i\tan\delta}$, where $n$ is the index of refraction and $\tan\delta$ is the loss tangent\cite{Essinger13}.

The relative optical axis orientation of the first and second layers is $\phi=54^\circ$. The optical axes of the first and third layers have the same orientation. The thickness of each birefringent layer of the AHWP is optimized such that the light path-length difference for the two crystal axes corresponds to a half wavelength at the central frequency of 120~GHz, and the stack can efficiently cover the two bands centered at 90~GHz and 150~GHz according to the Pancharatnam
model\textcolor{red}{\cite{Pancharatnam55}}.

\begin{table}
\centering
\begin{tabular}{ |c|c|c|c| } 
  \hline
	 Layer & Thickness [mm] & Index of refraction & Loss tangent [$10^{-4}$] \Tstrut\\
  \hline
  AR outer layer & 0.38 & 1.55 &   0.50 \Tstrut \\ 
  \hline
  AR inner layer & 0.27 & 2.52 &  56.5 \\ 
  \hline
  \multirow{2}{*}{HWP layer} &\multirow{2}{*}{3.75}	&3.05 ($n_o$)	&0.02 ($\tan\delta_o$)\\
  &			&3.38 ($n_e$)	&0.01 ($\tan\delta_e$)\\
  \hline
\end{tabular}
\caption{\label{tab:AHWP}
Design of the AHWP used in this work. The design matches the \textsc{Polarbear}-2 ambient-temperature AHWP design~\cite{Hill:2016jhd}. The position of both the AR layers is symmetric with respect to the three HWP layers. The values reported in this table are the central values of the corresponding measured quantities of the \textsc{Polarbear}-2 HWP~\cite{Hill:2016jhd}. As noted in main text, the exact configuration of the final SO HWP may differ from this configuration. $n_{o,e}$ are the indices of refraction along the ordinary and extraordinary axes of the HWP. The loss tangent is in general a function of frequency and temperature~\cite{parshin_dielectric_1994,Hill:2016jhd,savini_achromatic_2006}. In this analysis, we have assumed a constant loss tangent with frequency along the two axes for the sake of simplicity, and use values for a 50~K HWP~\cite{parshin_silicon_1995}. }
\end{table}

The Mueller matrix for the single-layer HWP takes the generic form of Eq.~\ref{eq:Mueller}, which allows for leakage into off-diagonal terms\cite{Bryan:10}. In the case of the AHWP, Eq.~\ref{eq:Mueller} takes a different form, i.e. it has more nonzero terms, to account for the fact that the AHWP is a stack of several layers rotated by a certain angle with respect to the first layer. We calculate a realistic Mueller matrix model of the SO HWP, $M^{\mathrm{HWP},k}$, as a product of the matrices corresponding to the three-sapphire stack AHWP and the AR coating layers.




\subsection{Variation with incident angle and frequency}\label{sec:hwpnutheta}
    \label{sec:HWP_vs_angle_and_nu}

Sapphire is a positive uniaxial crystal, with ordinary and extraordinary indices of refraction $n_o$ and $n_e$~\cite{Essinger13}. 
If an electromagnetic wave of frequency $\nu$ is transmitted from a medium with index of refraction $n_1$ to a sapphire crystal with incident angle $\theta$, the two components in which the linearly-polarized wave can be decomposed --- s- or TE and p- or TM waves --- are mixed inside the crystal to form the ordinary and extraordinary waves. The two components travel at different speeds through a uniaxial crystal. 
This induces a phase difference between the two components of the EM wave.

The effect induced by the HWP stack on the incoming wave depends on both the frequency of the wave and the incidence angle. In particular, at fixed $\theta$, the design of a single layer HWP can provide a phase difference of $\pi$ for a single frequency $\nu_c$ only. For frequencies that differ from $\nu_c$, the retarder does not behave as an ideal HWP. 

\begin{figure}
\begin{center}
\includegraphics[width=1.\textwidth]{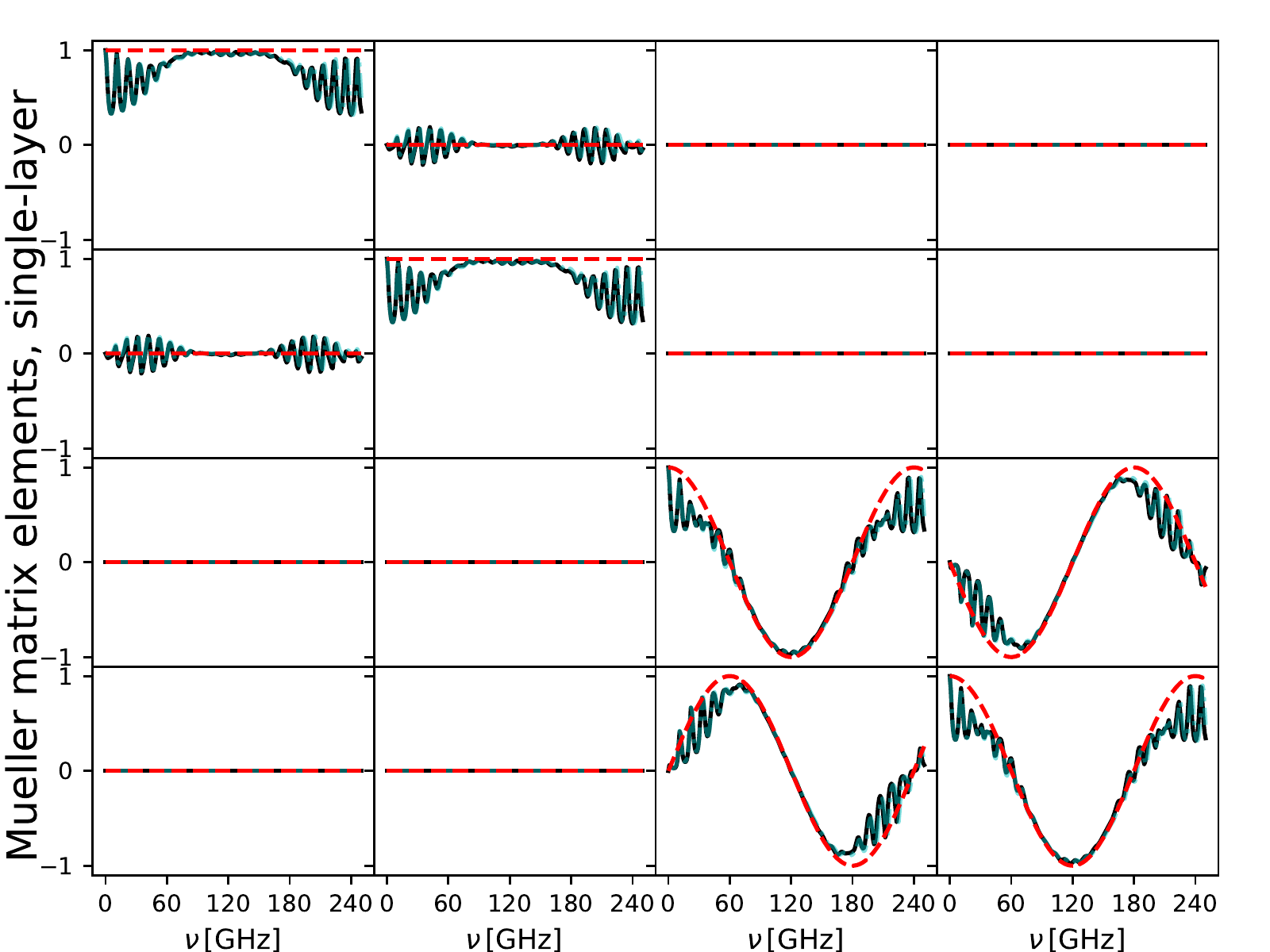}
\end{center}
\caption{Mueller matrix elements of a single-layer HWP as a function of the frequency $\nu$ of the incident wave. The design is that of the individual layers in a \textsc{Polarbear}-2-like (PB2) AHWP. The rotation angle is fixed and equal to $\chi=0^{\circ}$. The red dashed lines correspond to an ideal retarder. 
The black and cyan lines correspond to the simulations based on the PB2 HWP design and computed with the TMM. The black dashed lines are for $\theta=0^\circ$, whereas the cyan solid lines are for $\theta=20^\circ$. 
}\label{fig:Mij_one}
\end{figure}

\begin{figure}
\begin{center}
\includegraphics[width=1.\textwidth]{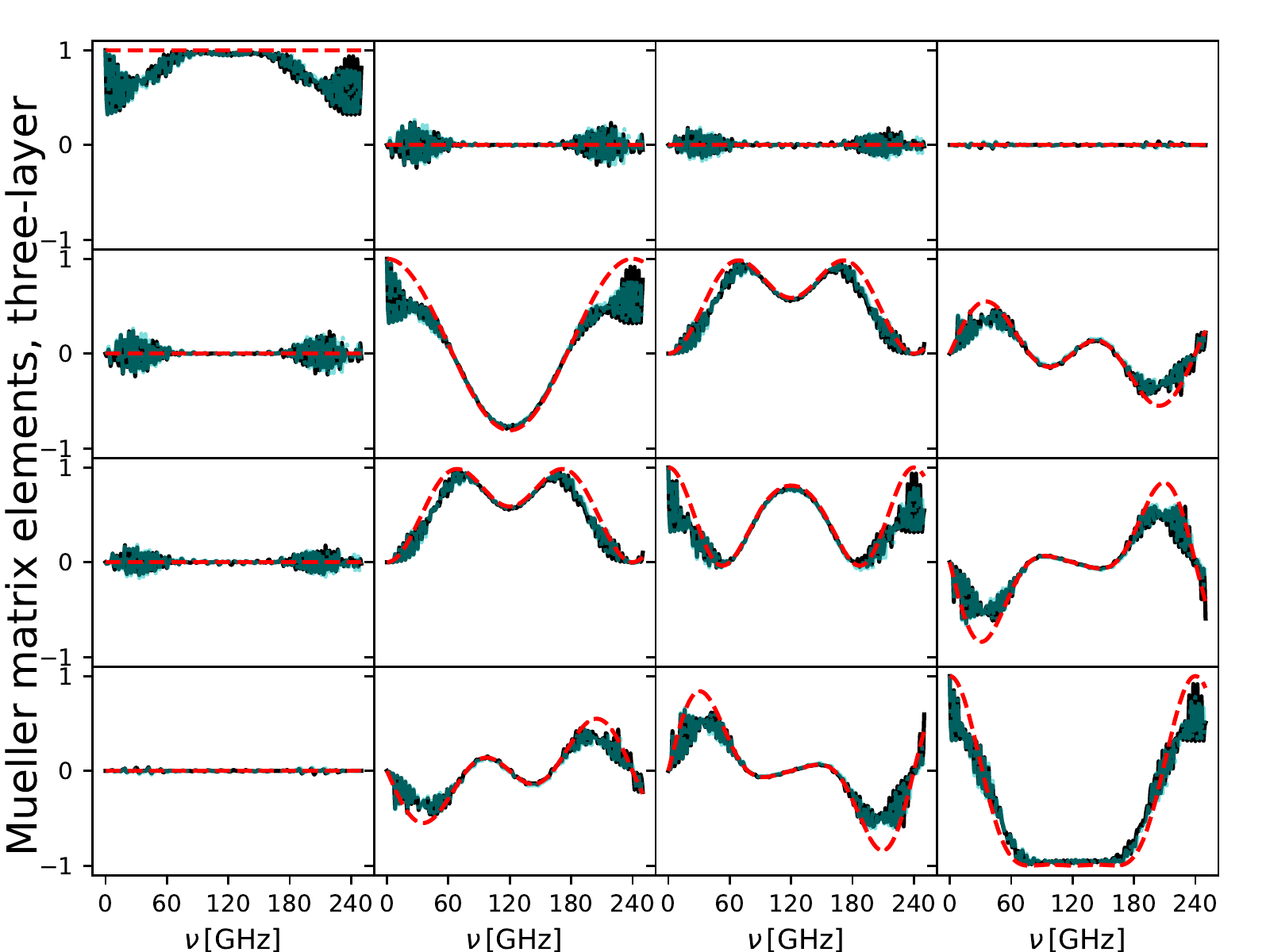}
\end{center}
\caption{Mueller matrix elements of a three-layer \textsc{Polarbear}-2-like AHWP as a function of the frequency $\nu$ of the incident wave in the same style as Fig.~\ref{fig:Mij_one}. If compared with Fig.~\ref{fig:Mij_one}, the different behavior of the Mueller matrix elements with the incident frequency is clearly visible in the case of a three-layer AHWP. In particular, the number of non-zero components is reduced with respect to the single-layer HWP. The non-zero components also exhibit a more complicated dependence from the incident frequency than in the single-layer case. The main reason for these clearly visible differences resides in the additional optical action induced by the multi-layer stacking with respect to the single layer HWP.
}\label{fig:Mij_three}
\end{figure}

Figures~\ref{fig:Mij_one} and~\ref{fig:Mij_three} depict the elements $M^{\mathrm{HWP},T}_{ij}$ for $i,j=1,2,3,4$ of a single-layer and three-layer Mueller matrix as a function of the frequency of the incoming wave and the incident angles $\theta=0^\circ,20^\circ$.  On the SAT, the aperture stop sets the field of view (FOV) of $34^{\circ}$, which gives a maximum incident angle of $17^{\circ}$ onto the HWP. We take $20^{\circ}$ as a conservative maximum incident angle to account for possible minor modifications to the SAT optical design. The elements in an ideal HWP with its thickness optimized for 120\,GHz and with zero reflection at boundaries (i.e., perfect AR coatings) are shown as red dashed lines. It is clear that the elements behave as an ideal HWP -- i.e. the phase difference is equal to $\pi$ --- only at the central frequency $\nu_0\simeq 120$~GHz. The dashed black line and the solid cyan line correspond to a design that resembles the \textsc{Polarbear}-2 HWP~\cite{Hill:2016jhd}, and are modeled through the TMM. The agreement with the ideal HWP is good for frequencies from 60~GHz to 180~GHz. Deviations from the ideal case are primarily due to differential properties of the sapphire and absorption inside the material, 
which is modeled through the loss tangent. The visible wiggles at all frequencies are primarily induced by multiple reflections between the boundaries of the dielectric layers, which are suppressed here by the presence of an AR coating. 
The solid black line corresponds to $\theta=0^\circ$, while the dotted-dashed cyan line corresponds to $\theta=20^\circ$. 
These figures show a small difference in the HWP behavior at $\theta=0^\circ$ and $\theta=20^\circ$ incidence angle. 
However, even a small difference can result in a significant effect if coupled to an incoming signal several orders of magnitude larger than the target signal (Sec.\,\ref{HWPSS_Intro}).

\begin{figure}
\begin{center}
\includegraphics[width=1.\textwidth]{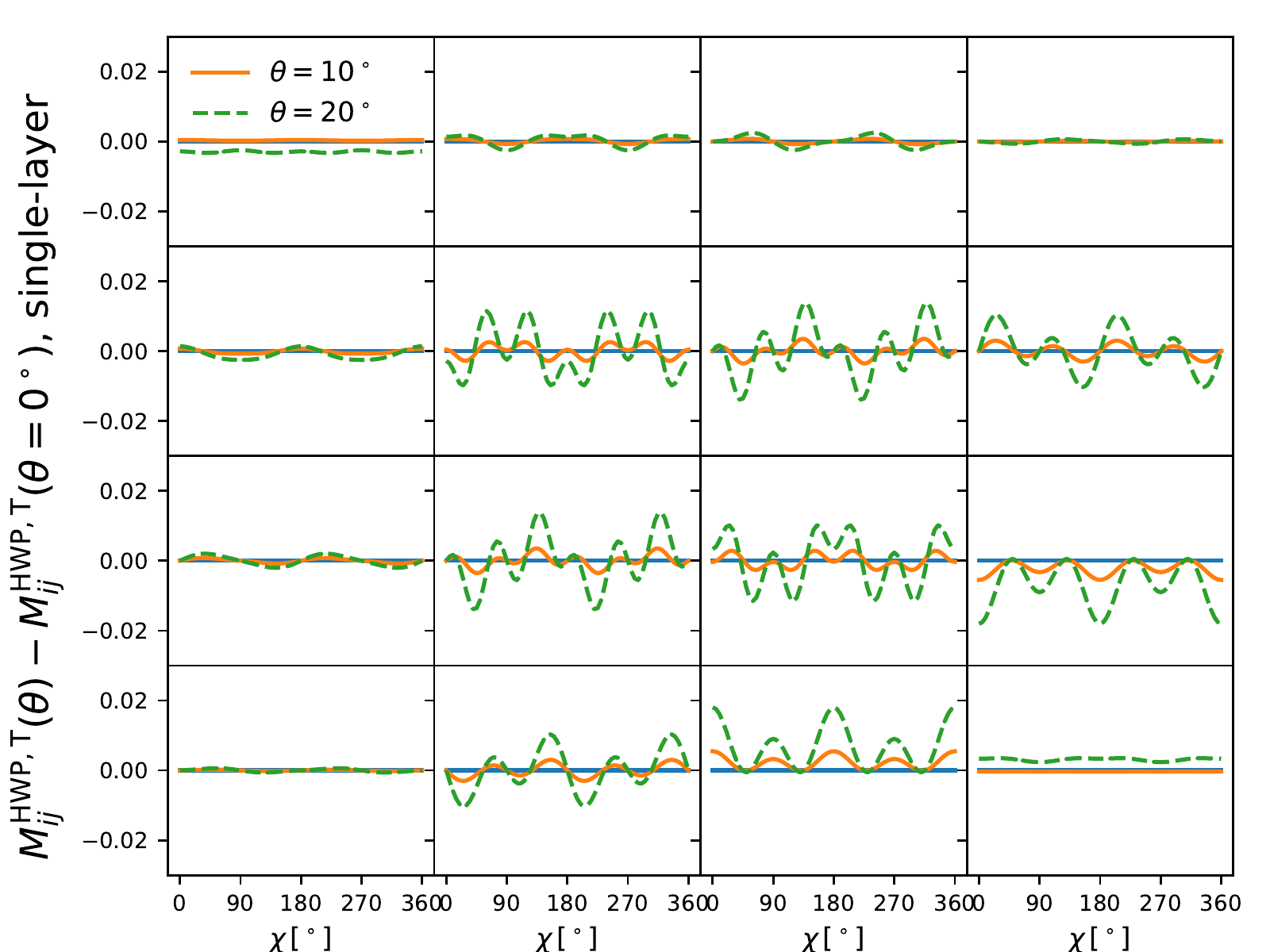}
\end{center}
\caption{Difference between the Mueller matrix elements for $\theta=10^\circ\,\mathrm{orange,\,solid},\,20^\circ\,\mathrm{green,\,dashed}$ and $\theta=0^\circ$, of a single-layer HWP as a function of the rotation angle $\chi$. The frequency is fixed and equal to $\nu_0=120\,\mathrm{GHz}$. The figure shows that the difference between non-normal incidence and normal incidence is very small. However, as it is discussed in Sec.~\ref{HWPSS_Intro}, it is not negligible.}\label{fig:Mij_diff_one}
\end{figure}

\begin{figure}
\begin{center}
\includegraphics[width=1.\textwidth]{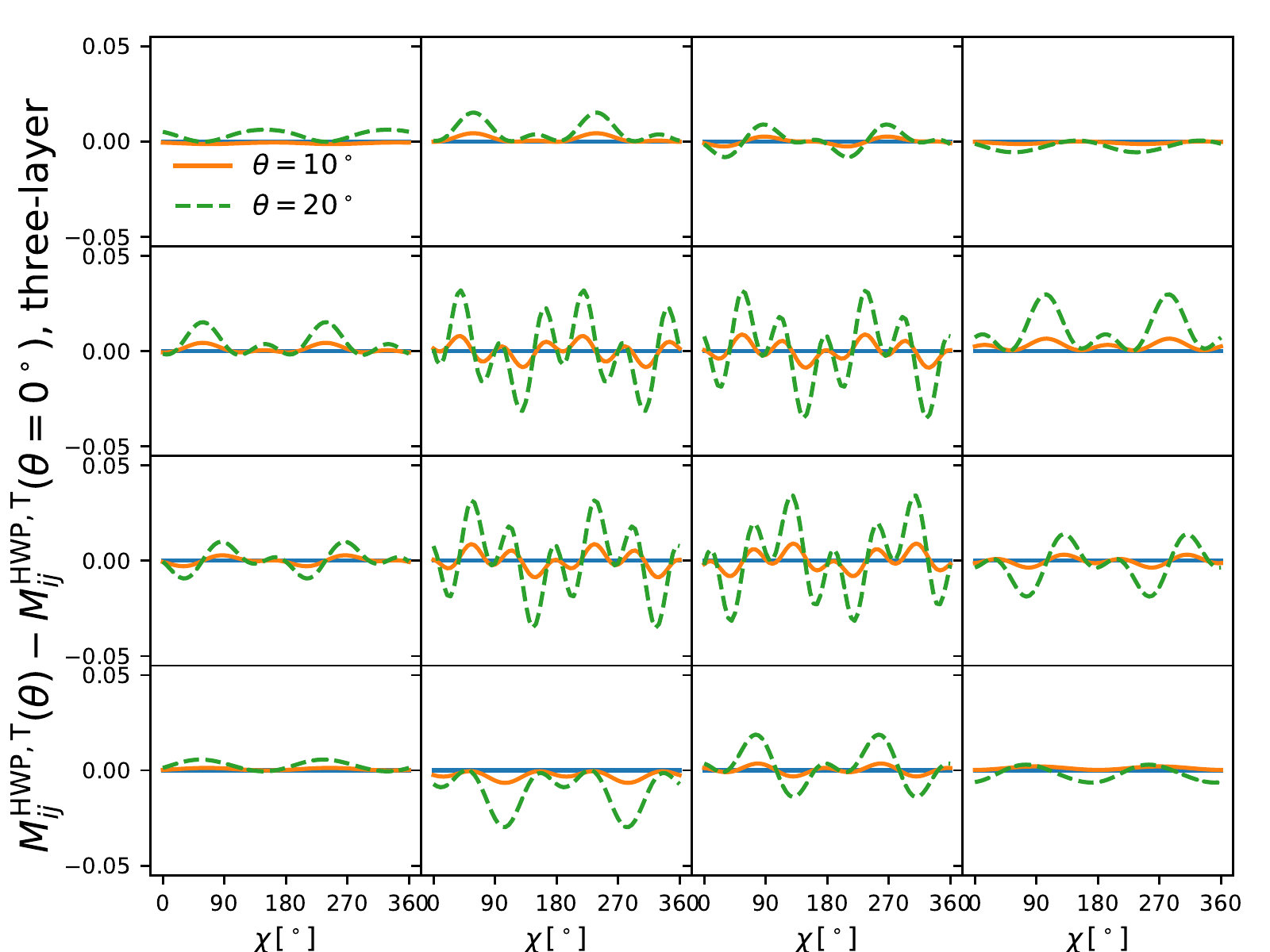}\\ 
\end{center}
\caption{Difference between the Mueller matrix elements for a three-layer AHWP in the same style as Fig.~\ref{fig:Mij_diff_one}. The frequency is fixed and equal to $\nu_0=90\,\mathrm{GHz}$.}\label{fig:Mij_diff_three90}
\end{figure}

\begin{figure}
\begin{center}
\includegraphics[width=1.\textwidth]{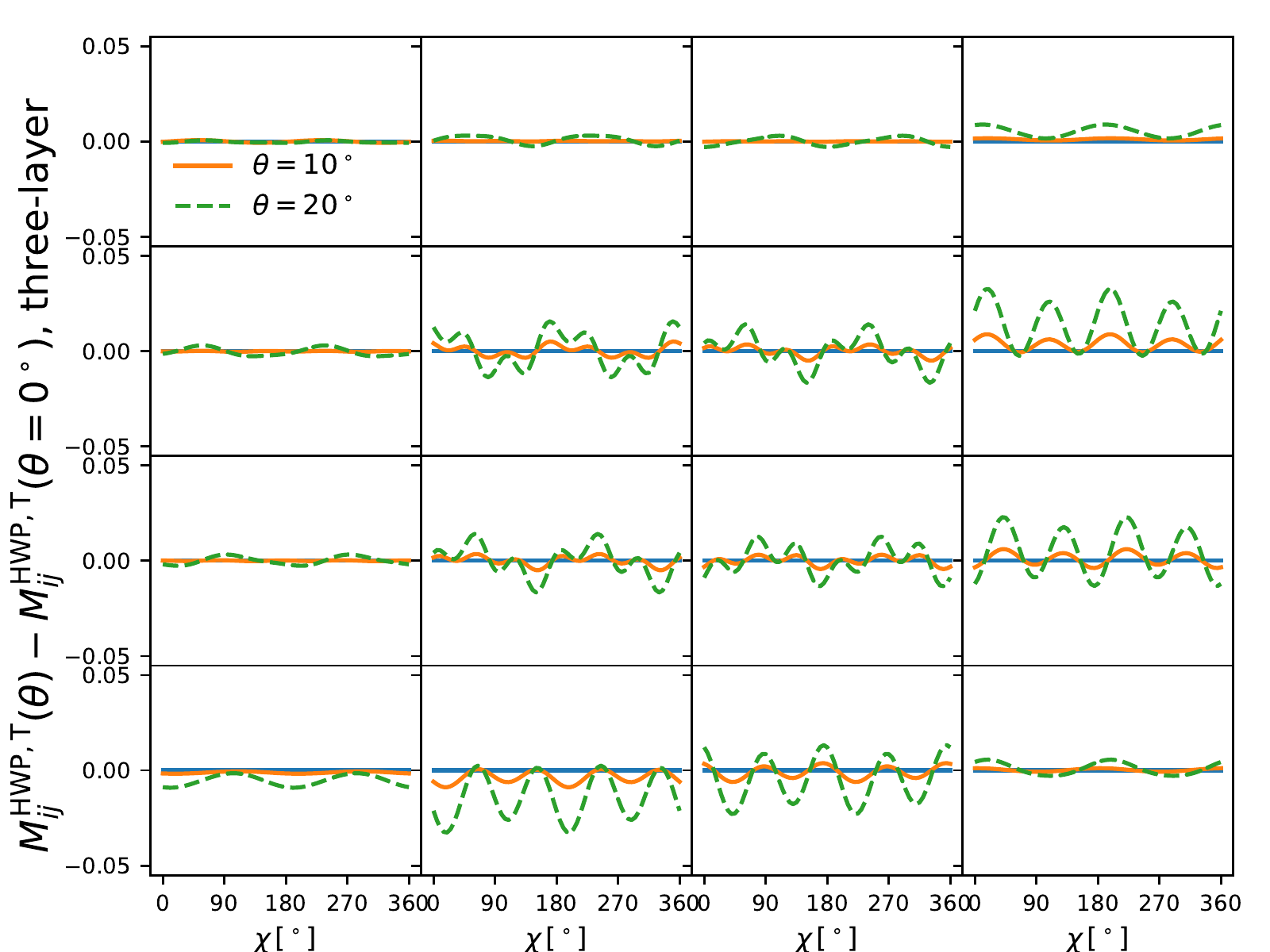}
\end{center}
\caption{Difference between the Mueller matrix elements for a three-layer AHWP in the same style as Fig.~\ref{fig:Mij_diff_one}. The frequency is fixed and equal to $\nu_0=150\,\mathrm{GHz}$.}\label{fig:Mij_diff_three150}
\end{figure}

Figures~\ref{fig:Mij_diff_one}, \ref{fig:Mij_diff_three90} and~\ref{fig:Mij_diff_three150} show the difference $M^{\mathrm{HWP},T}_{ij}(\theta)-M^{\mathrm{HWP},T}_{ij}(\theta=0^{\circ})$ as a function of the rotation angle $\chi$ for a fixed frequency $\nu_0=120\,\mathrm{GHz}$, for a single-layer HWP and a three-layer AHWP, respectively. For the AHWP, the difference with respect to the $\theta=0^{\circ}$ case is always $<0.05$\footnote{We note that the AR coating plays a major role in keeping this difference low. In the absence of the AR coating, the difference can be a factor of a few higher.}. However, in Sec.~\ref{HWPSS_Intro}, we will see that this small difference can induce non-negligible effects.

\subsection{Differential Absorption and Emission}\label{sec:absorption}
    \label{sec:Differential_Absorption}

The HWP emission can be described with the Mueller matrix of a polarizer with $p_{e}$ and $p_{o}$ being the emissivities along the extraordinary and ordinary axis, respectively. They correspond to the $p_x$ and $p_y$ elements of a Mueller matrix polarizer\cite{Collett93}. 
The emissivity along each axis can be estimated with the radiative transfer in the case of an optically thin medium: 
\begin{equation}
\label{eq:HWP_emissivity}
p_{o,e}= 1 -\exp{(- \alpha_{o,e}\,d/\cos{\theta_{o,e}})},
\end{equation}
where $\alpha_{o,e}$ is the absorption coefficient along the two axes at 50~K, $d$ is the HWP thickness and $\theta_{o,e}$ is the refraction angle along the two axes. The absorption coefficient can be written in terms of the loss tangent $\tan (\delta)$ as: 
\begin{equation}
\label{eq:absorption coefficient}
\alpha_{o,e}=\tan (\delta)2\pi n_{o,e}\nu / c
\end{equation}
with $c$ the speed of light\cite{Lamb1996}. The differential absorption is equal to the differential emissivity. 
We use loss tangents given in Table \ref{tab:AHWP} which were measured by Parshin\cite{parshin_silicon_1995} at 50~K.
This suppresses the absorption coefficient by a factor of 100 compared to sapphire at room temperature.

\section{Estimating the HWP Synchronous Signal}\label{sec:HWPSS_model}
\label{HWPSS_Intro}

In this section, we assume a realistic optical configuration of SO SAT (Sec.~\ref{sec:sac_hwp}), and calculate the HWPSS for the setup combining the Mueller matrix of the optical chain with the Mueller matrix of HWP described in the previous section (Sec.~\ref{sec:hwpnutheta}). We then provide the interpretation of the dominant sources of the HWPSS in Sec.~\ref{sec:hwp_hwpss} and~\ref{sec:hwp_updown}.

During observations, 
each detector sees a modulated signal $d_m$ as a function of the HWP rotation angle $\chi$:
\begin{equation}
\label{eq:demod_signal}
d_m(\chi) = I + \text{Re}\left\{\varepsilon_\text{pol} e^{4 i \chi}
(Q + i U)
\right\}
+ A(\chi) + \mathcal{N}_\mathcal{W},
\end{equation}
where $I$, $Q$, and $U$ are the components of the incoming Stokes vector $S_\mathrm{in}=(I,Q,U,V)$\footnote{We assume here that the circularly polarized component is negligible\cite{Nagy17}.}, $\varepsilon_\text{pol}$ is the polarization efficiency,  $A(\chi)$ is a spurious signal called the HWPSS, and $\mathcal{N}_\mathcal{W}$ is the white noise. As we will see in more detail in this section, there are various sources of HWPSS.

The HWPSS $A(\chi)$ defined in Eq.~\ref{eq:demod_signal} can be modeled using a series of harmonics of the HWP rotation frequency $f$. 
The amplitude of each harmonic can be split into a term $A_n$ that is constant in time throughout the scan assuming a stationary instrument,
and a term that is proportional to the incoming intensity, with proportionality constant $a_n^\text{opt}$.
The HWPSS can then be written as:
\begin{equation}
\label{eq:HWPSS_def}
A(\chi) = \sum_{n = 1}^\infty (A_n + a_n^\text{opt} I) \cos(n \chi + \phi_n).
\end{equation}

Signals modulated at 2$f$ and 4$f$ can be generated in a number of ways. Here, the signal at 4$f$ is of particular interest since it is at the center of the science band. The HWP itself can have differential transmission, reflection (Sec.~\ref{sec:Mueller}), and emission (Sec.~\ref{sec:absorption}) along the ordinary and extraordinary axes generating a 2$f$ signal.
Irregularities in the HWP AR coating or inherent $\chi$ dependence in the HWP Mueller matrix (e.g. Fig.~\ref{fig:Mij_three}) caused by  non-normal incidence angles can create 2$f$, 4$f$, and higher order harmonic signals.
Optical elements upstream and downstream of the HWP generate polarized light through differential transmission, reflection, and emission that can be modulated by the HWP creating a 4$f$ signal. In the presence of detector nonlinearity, both the 2$f$ and 4$f$ components of the HWPSS contaminate the polarization band and are sources of systematic error.

In this section, we will describe these sources of spurious $2f$ and $4f$ signals and how accurately we are able to model them for the SAT optical chain shown in Table~\ref{tab:optical_chain}.
The coefficients $A_2$, $A_4$, $a_2^\text{opt}$, and $a_4^\text{opt}$ that we calculate are given in Table~\ref{tab:HWPSS_coeffs}. Experimentally, we will be able to characterize this polarization leakage by making polarization maps for known unpolarized sources such as Jupiter. 
Note that the optical chain and HWP design used in this calculation are not finalized, so the results in this section are not final.
Our results demonstrate our ability to model these systematic effects for an SO-like system and highlight which elements of our system we expect to dominate the HWPSS. 


\begin{table}
\centering
\begin{tabular}{ |c||c|c|c|c| } 
  \hline
	& $A_2$ (mK$_\text{CMB}$) & $a_2^\text{opt}$ (\%) & $A_4$ (mK$_\text{CMB}$) & $a_4^\text{opt}$ (\%) \Tstrut\\
  \hline
  95 GHz ($0^\circ$) & 210.77 & 0.87 &  0.00 & 0.00 \Tstrut \\ 
  95 GHz ($20^\circ$) & 214.39 & 0.87 & 185.13 & 0.76 \\ 
  145 GHz ($0^\circ$) & 197.76 & 0.65 & 0.00 & 0.00\\ 
  145 GHz ($20^\circ$) & 220.61 & 0.72 & 178.57 & 0.63\\ 
  \hline
\end{tabular}
\caption{\label{tab:HWPSS_coeffs}
The 2$f$ and 4$f$ HWPSS coefficients for the middle frequency (MF) band at 0$^\circ$ and 20$^\circ$ incident angles estimated with the method described in Sec.~\ref{HWPSS_Intro}.
}
\end{table}

\begin{table}[b]
\centering
\begin{tabular}{ |c|c|c|c|c| } 
  \hline
	$\#$	&Element & Temp (K) &$T_{s-p}$ (90 GHz) & $T_{s-p}$ (145 GHz) \Tstrut\\
    \hline
    1  &Window	 &	273 & 0.047 \% & 0.094 \%\\
    2 & DSIR Filter &	273 & 0 & 0 \\
    3 & DSIR Filter &	220 & 0 & 0 \\
    4 & Alumina Filter &	60 & 0.329 \%  & 0.367 \%\\
    5 & Sapphire HWP & 40 & -------- & -------- \\
    6 & Alumina Filter &	4  & 0.329 \% & 0.367 \%\\
	7 &  Lyot Stop &	2 & 0 & 0 \\
	8 &  Silicon Lenses (3) &	1.5 & 0& 0 \\
	9 &  Low Pass Filters (2) &	1.5 & 0& 0 \\
	10 &  Low Pass Filter &	0.1 & 0& 0 \\
    \hline
\end{tabular}
\caption{\label{tab:optical_chain}
Optical chain for the SAT. For each element, the corresponding temperature stage is reported. The numeration runs from the outermost element (the window at $T=273$~K) to the innermost elements (the low pass filters at $T<1$~K). The sapphire HWP is located between alumina filters and is kept at a cryogenic temperature of $40$~K.
The differential transmission coefficients $T_{s-p}$ are calculated at
a $20^\circ$ incidence angle.
}
\end{table}

\subsection{The SAT Cryogenic HWP}\label{sec:sac_hwp}
    \begin{figure}
\begin{center}
\includegraphics[width=0.7\textwidth]{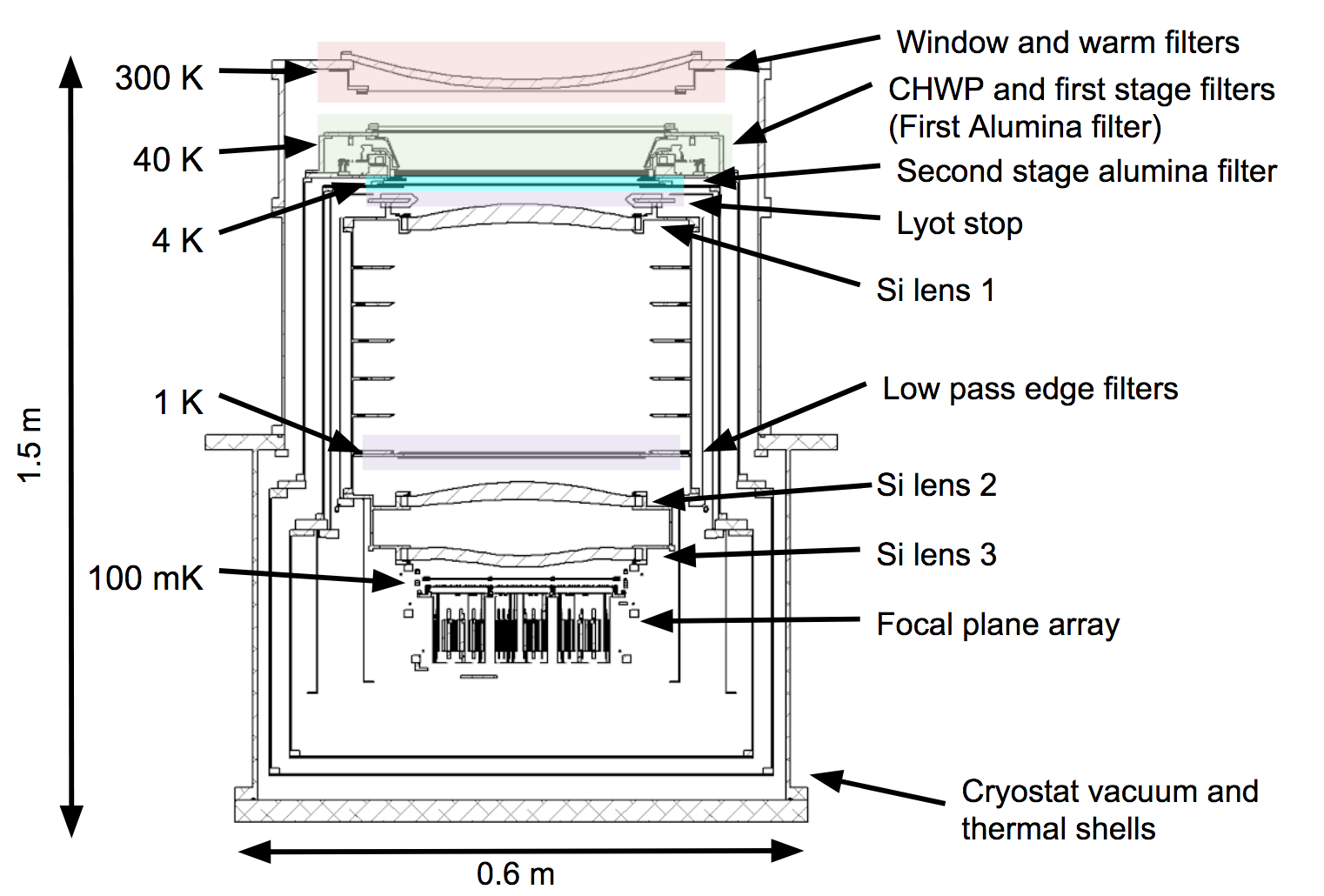}
\end{center}
\caption{Optical chain of the SAT and the planned temperature stages.
The HWP lies at the 40~K stage. 
}
\label{fig:SAT_optical_chain}
\end{figure}

The cryogenic CRHWP for the SAT will be located at the 40~K temperature stage\footnote{For our aim there is not significant difference in considering the HWP loss tangent at 40 or 50K, as considered in Tab.\,\ref{tab:AHWP}.}. It will be rotated by superconducting magnetic bearings~\cite{Hanany03b, Hill18}.
Due to current design uncertainty, the temperatures and the exact details of each optical element are not finalized, but the current state of the optical chain and temperatures are shown in Table \ref{tab:optical_chain} and Figure \ref{fig:SAT_optical_chain}.



Several considerations were taken into account when choosing the location of the HWP. Any $I\rightarrow P$ leakage in front of the HWP will be modulated into $A_{4}$, so a HWP should ideally be as far sky-side as possible. 
However, in order to minimize HWP thermal loading, i.e. its thermal emission, we choose to adopt cryogenic HWP and thus place the HWP behind the cryostat window and between alumina filters. Additionally, the HWP lies close to the aperture stop, where the detector beams intersect, so we do not need to worry about spatial non-uniformity of the HWP as seen by different detectors. Placing the HWP at the 4-K stage would have further reduced thermal loading, but this would have also required more optical elements in front of the HWP, increasing $I \rightarrow P$. Taking these considerations into account, the reduction in the HWP emission from 40~K down to 4~K would not have resulted in significant gain when compared with more challenging HWP operation at 4~K and possible systematics due to the increased $I\rightarrow P$ leakage.


\subsection{Light Propagation and Estimation of the HWPSS}
    \label{sec:Light_Propagation_and_Estimation_of_HWPSS}

To calculate the harmonic amplitudes in Eq.~\ref{eq:HWPSS_def}, we must be able to calculate the Stokes vector incident on both the sky- and detector-side of the HWP. To do this, we use the transmission, reflection, and absorption coefficients along the $s$ and $p$ polarization directions for each optical element to propagate light throughout the telescope. The $s$ and $p$ directions are defined to be orthogonal to the plane of incidence, and parallel to the plane of incidence respectively.
We assume that the light is angled such that these directions correspond to the $y$ and $x$ directions.

For the window and alumina filters we calculate the coefficients as a function of incident angle using the TMM. We found other elements are significantly less relevant in our analysis, and thus are showing results with their effects set zero for simplicity.
For the TMM, we require the thickness and complex index of refraction of the optical element and of the AR coating.
The coefficients are dependent on the incident angle of the incoming light. 

For our conservative estimation, we make a few simplifications and assumptions.
We treat each optical element as an infinite horizontal slab, and do not take into account curvature of the element or walls of the telescope. 
 We assume that the beam from a single detector is perfectly collimated after the silicon lenses, and so all light travels through the main optics at a single incidence angle. For the SAT, it is a good approximation to assume that the incident angle is roughly proportional to the distance of the detector from the center of the array, with $\theta = 17^\circ$ for a detector at the edge and $\theta=0^\circ$ for a detector at the center.
We also assume 100\% beam coupling and detector efficiency.
A more detailed simulation would use a more accurate beam coupling and detector efficiency, and would use a SAT ray-tracing to get more accurate incidence angles.

The amount of polarized power created through differential transmission, reflection, or emission is determined by the differential coefficients
\begin{equation}
T_{s - p} = \frac{T_s - T_p}{2}, \qquad
R_{s - p} = \frac{R_s - R_p}{2}, \qquad \text{and} \qquad
A_{s - p} = \frac{A_s - A_p}{2}.
\end{equation}

We use an iterative process to calculate the
forward and backwards traveling Stokes vectors $S_\text{fw}^\text{HWP}$ and $S_\text{bw}^\text{HWP}$ which are incident on the sky-side and detector-side of the HWP, respectively.
We propagate light through each element, calculating the forward and backward propagating Stokes vectors on the $n^\text{th}$ element as:
\begin{equation}
\label{eq:stokes def}
S_\text{fw}^n = 
S_\text{emit}^{n-1} + M_\text{trans}^{n-1} S_\text{fw}^{n-1} + M_\text{refl}^{n-1} S_\text{bw}^{n-1}
\qquad \text{and} \qquad
S_\text{bw}^n = 
S_\text{emit}^{n+1} + M_\text{trans}^{n+1} S_\text{fw}^{n+1} + M_\text{refl}^{n+1} S_\text{fw}^{n+1}.
\end{equation}
The Mueller matrices are determined by the transmission and reflection coefficient of the optical elements. 
These are given by:
\begin{equation}
\label{eq:Optical element mueller matrix}
M_\text{trans} = 
\left(
  \begin{array}{cccc}
    T_\text{s+p} & T_\text{s-p} & 0 & 0 \\
    T_\text{s-p} & T_\text{s+p} & 0 & 0 \\
    0 & 0 & 1 & 0 \\
    0 & 0 & 0 & 1 \\
  \end{array}
\right)
\qquad \text{and} \qquad
M_\text{refl} = 
\left(
  \begin{array}{cccc}
    R_\text{s+p} & R_\text{s-p} & 0 & 0 \\
    R_\text{s-p} & R_\text{s+p} & 0 & 0 \\
    0 & 0 & 1 & 0 \\
    0 & 0 & 0 & 1 \\
  \end{array}
\right).
\end{equation}
By iterating the propagation process multiple times, using the incident Stokes vector from the previous iteration to calculate the detector-side Stokes vectors, we take into account light that could reach the detector after multiple reflections.

Once we have the incident Stokes vectors on the HWP, we can compute the modulated Stokes vector seen by the detector as a function of the HWP rotation angle
\begin{equation}
\label{eq:mod_computation}
S_\text{det} = \varepsilon M^\text{Pol}  
(M^\mathrm{stack}_\text{trans}(\chi) S^\text{HWP}_\text{fw} + 
M^\mathrm{stack}_\text{refl}(\chi) S^\text{HWP}_\text{bw}+S_\text{emit}^\text{HWP}(\chi)),
\end{equation}
where $M^{\mathrm{Pol}}$, the Mueller matrix of a linear polarizer with $p_x=1$ and $p_y=0$, 
\begin{equation}\label{eq:Mpol}
M^{\mathrm{Pol}}=\frac{1}{2}\left(
  \begin{array}{cccc}
    p_x^2+p_y^2 & p_x^2-p_y^2 & 0 & 0 \\
    p_x^2-p_y^2 & p_x^2+p_y^2  & 0 & 0 \\
    0 & 0 & 2p_xp_y & 0 \\
    0 & 0 & 0 & 2p_xp_y \\
  \end{array}
\right)
\end{equation}
represents a polarization sensitive detector,
and $\varepsilon$ is the cumulative transmission efficiency between the HWP and the detector. $S_\text{emit}^\text{HWP}(\chi)$ is the term due to the HWP emissivity (Sec.\,\ref{sec:hwp_hwpss}). The modulated signal from the detector $d_m$ is the intensity of $S_\text{det}$.
The HWP transmission and reflection Mueller matrices are both calculated using TMM, as described in Section\,\ref{sec:sapphire_mueller}. $M^\mathrm{stack}_\text{trans}(\chi)$ and $M^\mathrm{stack}_\text{refl}(\chi) $, respectively, are both the Mueller matrices of the AHWP in transmission and in reflection versus the HWP rotation angle $\chi$.
We fit this signal with the amplitudes and phases of the first eight harmonics 
from the intensity independent part of Eq.~\ref{eq:HWPSS_def} to determine the coefficients $A_n$.
\subsection{HWPSS contributions from the HWP}\label{sec:hwp_hwpss}
    \label{sec:HWPSS_from_HWP}

Most polarized signal generated by the HWP itself will rotate at the same frequency as the HWP, coupling to the detector at $2f$. However, a small amount can be modulated up to the $4f$ band or higher harmonics either from non-normal incidence angles or irregularities in the HWP AR coating (see e.g. examples of non-trivial behavior of an AHWP in Section \ref{sec:HWP_vs_angle_and_nu}).

Since the HWP Mueller matrix used in Eq.~\ref{eq:mod_computation} is calculated using the generalized TMM described in Sec.~\ref{sec:sapphire_mueller}, 
the harmonic amplitudes we fit already have the non-normal incidence angle taken into account. However, any HWPSS emitted by the HWP is not taken into account by the fitting and must be added separately as $S^\text{HWP}(\chi)$ in Eq.\,\ref{eq:mod_computation}. The emissivities along the two axes are calculated using Eq.~\ref{eq:HWP_emissivity}.

For the optical chain defined in Tab.~\ref{tab:optical_chain}, we calculate the fraction of the $A_2$ signal that is generated by the transmission, reflection, and emission of the HWP, corresponding to the three terms in Eq.\,\ref{eq:mod_computation}. The results are quoted in Table~\ref{tab:A2_division}. As we would expect, the $A_2$ signal is dominated by differential transmission through the HWP, and the polarized emission is small because the HWP is cooled to 40~K. 

The contribution to the $A_4$ component due to the modulation of unpolarized light by transmission and reflection by the HWP at large incidence angles is given in Table~\ref{tab:A4_division}.


\begin{table}
\centering
\begin{tabular}{ |c||c|c|c|c| } 
  \hline
	& Trans (mK$_\text{CMB}$) & Refl (mK$_\text{cmb}$)& Emit (mK$_\text{CMB}$)& Total (mK$_\text{CMB}$)\\
  \hline
  95 GHz ($0^\circ$) & 192.76 & 16.42 &  1.58 & 210.77\\ 
  95 GHz ($20^\circ$) & 195.48 & 17.32 & 1.58 & 214.39\\ 
  145 GHz ($0^\circ$) & 172.91& 21.73 & 3.12 & 197.76 \\ 
  145 GHz ($20^\circ$) & 192.78 & 24.70 & 3.12 & 220.61\\ 
  \hline
\end{tabular}
\caption{\label{tab:A2_division}
Contribution to the $A_2$ amplitude (in $\mathrm{mK_\mathrm{cmb}}$) in the $A(\chi)$ HWP synchronous signal from transmission, reflection, and emission for the 90/150~GHz band. The amplitudes are reported for two extreme values of the incident angle $\theta$: normal incidence where $\theta=0^\circ$ and the most extreme slant incidence case of $\theta=20^\circ$.
}
\end{table}

\subsection{HWPSS contributions from optics upstream and downstream of the HWP}\label{sec:hwp_updown}
    \label{sec:HWPSS_upstream_and_downstream}

Polarized light generated by optical elements upstream of the HWP will be modulated at $4f$ when transmitted by the HWP.
The majority of this polarized light is generated through the differential transmission or reflection of unpolarized light by only a few optical elements at large incidence angles.
The unpolarized power incident on the optical elements is primarily created by the  atmosphere and the window. 
We utilize atmospheric simulations of the observation site generated by the AM atmospheric modeling code, using a PWV of 1 mm and an elevation of 50$^\circ$\footnote{\url{https://www.cfa.harvard.edu/~spain/am/}.},
for which the total power emitted is about 12~K$_\text{CMB}$ in the MF band.
For the window element we are assuming an absorption coefficient of $1\%$ for 150 GHz and $0.5\%$ for 90 GHz, and we assume a $1.5\%$ spillover of the beam onto a warm surface.
The spillover and window emission together add a total power of 11~K$_\text{CMB}$ at 150 GHz and 6~K$_\text{CMB}$ at 90 GHz to the system.

Only the alumina filters and the window have 
non-negligible differential transmission and reflection coefficients.
The alumina filter has a thickness of 3~mm, and an index of refraction of $n_0 = 3.1 + 8 \times 10^{-5} i$.
The window is made of UHMWPE, and has a thickness of 1~cm and an index of refraction of $n_0 = 1.5 + 1 \times 10^{-4} i$.
For each element we use a 2-layer AR coating\cite{Rosen13}, where the outer and inner layers have indices of refraction $n'$ and $n''$ given by
\begin{equation}
\label{eq:AR_indices}
n' = n_0^{1/3} \qquad \text{and} \qquad
n'' = n_0^{2/3}.
\end{equation}
The corresponding thicknesses of the two layers are 
\begin{equation}
\label{eq:AR_thickness}
d' = \frac{\lambda_0}{4 n'} \qquad \text{and} \qquad 
d''= \frac{\lambda_0}{4 n''}.
\end{equation}
For the MF band, we use $\lambda_0 = 2.5$~mm.
Because the absorption is negligible, for the window and the filter we have 
$T_{s-p}\sim R_{s-p}$. 
The differential coefficients for the optical elements for a $20^\circ$ incidence angle are shown in Table \ref{tab:optical_chain}.
Even though the differential transmission and reflection coefficients are almost equal, the amount of  $4f$ signal caused by differential transmission is generally much larger than the amount caused by differential reflection because there is more light incident on the sky-side of the optical elements than the detector-side of the element.

A breakdown of the $A_4$ signal for a $20^\circ$ incidence angle can be seen in Table \ref{tab:A4_division}. The table shows how much of the $4f$ signal is generated by differential transmission and reflection from the window and the alumina filter on the sky-side of the HWP. For the HWP, we show how much $4f$ signal is generated from the transmission or reflection of unpolarized light at a large incidence angle.
We also take into account polarized light being reflected off of the detector side of the HWP, but this is mostly modulated at $2f$ and its contribution to $A_4$ is negligible.

We observe that the majority of the $4f$ signal comes from differential transmission through the alumina filter and the transmission of unpolarized light through the HWP at large non-normal incidence.
At 95~GHz, the amount of $4f$ signal created by the HWP is larger than the amount created by the alumina filter.
The $4f$ signal created by differential transmission through the window, and differential reflection off of the filter are less significant, but are still non-negligible.

\begin{table}
\centering
\begin{tabular}{ |c||c|c|c|c| } 
  \hline
  & \multicolumn{2}{|c|}{95 GHz} & \multicolumn{2}{|c|}{145 GHz} \\
  \hline
  & Trans (mK$_\text{CMB}$)& Refl (mK$_\text{CMB}$)& 
    Trans (mK$_\text{CMB}$)& Refl (mK$_\text{CMB}$)\\
  \hline
  Window & 6.47  & 2.02  & 13.11 & 6.81  \\
 Alumina Filter & 70.73 & 8.19 & 97.93 & 17.41 \\
  HWP    & 85.83 & 6.36  & 36.13 & 4.08  \\
  \hline
\end{tabular}
\caption{\label{tab:A4_division}
The breakdown of sources of $A_4$ signal for an incidence angle of $20^\circ$.
For the window and filter, the Trans and Refl columns are the magnitude of the polarized signal created through differential transmission and differential reflection.
For the HWP, the Trans and Refl columns are the magnitude of the $A_4$ signal generated by transmitting or reflecting light at a non-normal incidence angle.
}
\end{table}

\section{Polarization Leakage from Nonlinearity}\label{sec:NL}
    Another major source of $I\rightarrow P$ leakage is the non-linear behavior of the detector response~\cite{Salatino10}. This effect is caused primarily by the dependence of the detector responsivity and time constant on the temperature of the system and incoming loading. 
The dominant source of $I\rightarrow P$ leakage observed in experiments such as \textsc{Polarbear} \cite{Takakura:2017ddx} and EBEX is the coupling of incoming intensity to the HWPSS, creating leakage from the $0f$ (constant) band to the $4f$ band\footnote{A 4$f$ signal could be also created from a 2$f$ signal through detector non-linearity or saturation effects~\cite{Salatino10}, but this is not the main contribution in fielded experiments.}. 
We show that for the SAT the $I\rightarrow P$ leakage caused by non-linearity is not dominant, and that it is smaller than the leakage created by differential transmission through the alumina filter.

We use the non-linearity model described in the \textsc{Polarbear} analysis~\cite{Takakura:2017ddx} where the modified timestream is given by:






\begin{equation}
\label{eq:NL_def}
d'(t) = [ 1 + g_1 d(t) ] d(t - \tau_1 d(t)).
\end{equation}
$g_1$ and $\tau_1$ are the coefficients representing the lowest order non-linearity in gain and time constant, respectively.


To leading order, the coupling of the unpolarized $I(t)$ signal with the HWPSS induced by Eq.\,\ref{eq:NL_def} adds $I\rightarrow P$ leakage 
with proportionality constant
\begin{equation}
\label{eq:NL_leakage_coeff}
a_4^\text{NL} = (2 g_1  + i \omega_\text{mod} \tau_1 )A_4,
\end{equation}
where $\omega_\text{mod} = 4 (2 \pi f)$. 

\subsection{Simulating Nonlinearity $I \rightarrow P$}
	We analyze the non-linearity induced $I\rightarrow P$ leakage using time-domain simulations with the \texttt{s4cmb} systematics pipeline. The pipeline starts by generating a sky map to use as input based on fiducial bandpowers that include Gaussian realizations of lensed power spectra from CAMB\cite{camb}. 
Currently the pipeline simulates a focal plane consisting of $4$ pairs of detectors, with the paired detectors sensitive to orthogonal polarizations. 
The pipeline then runs a number of Constant Elevation Scans (CESs) determined by a given scan-strategy, which can be either wide- or deep-field.
For each CES, TOD is generated for each detector from the input sky. This TOD includes correlated $1/f$ noise and uncorrelated white noise. The simulation of correlated noise is based on realistic observations of atmospheric fluctuations. The level of white noise can be adjusted to roughly match the expected sensitivity given an experimental setup.

We inject the HWPSS into the CMB-plus-noise TOD of each detector using the parameters in Sec.~\ref{sec:HWPSS_model} and Tab.~\ref{tab:HWPSS_coeffs}. We then add non-linearity effects following Eq.~\ref{eq:NL_def}.
For each detector, $g_1$ and $\tau_1$ are selected from a normal distributions centered at $g_1=-4.2 \cdot 10^{-3} \rm K^{-1}$ and $\tau_1=5\cdot 10^{-5}$~sK$^{-1}$ with $\sigma(g_1)=2.5$\%$g_1$ and $\sigma(\tau_1)=2.5\%\tau_1$, in agreement with Takakura et al.~\cite{Takakura:2017ddx}.

To remove the intensity-independent portion of the HWPSS, 
we use the technique developed for the EBEX analysis pipeline~\cite{Didier17}.
This splits the TODs into hour-long chunks, and then fits and subtracts the first eight HWP harmonics to account for the steady $2f$ and $4f$ signals added along with any higher order harmonics created by the non-linearity.


Using the HWPSS-removed timestreams we can estimate the magnitude of the $I\rightarrow P$ leakage created by non-linearity for each detector by looking at the coupling of the polarization timestreams with atmospheric noise.
For detectors at the edge of the array with a $20^\circ$ incidence angle we see a total leakage coefficient of $a_4^\text{total} = 0.78\%$, with about $0.15\%$ coming from the addition of non-linearity, which is in agreement with Eq.~\ref{eq:NL_leakage_coeff} using our values for $g_1$ and $\tau_1$. 
For a detector with $0^\circ$ incidence angle, the $A_4$ component is 0, so there is no direct $0f\rightarrow 4f$ leakage~\cite{Essinger16}.
We observe the leakage magnitude to $a_4^\text{total} = 0.02\%$, 
which must mostly come from $2f\rightarrow 4f$ leakage.
We do not apply any $I\rightarrow P$ removal to the data, so these values represent upper limits.





\section{Conclusion and Future Work}\label{sec:conclusions}

The Simons Observatory will be one of the most sensitive CMB experiments when it deploys. Thus, it is necessary to understand the sources of systematic contamination and develop mitigation strategies to reach the SO's ambitious science goals. 
A continuously-rotating HWP is expected to benefit CMB polarization measurements by offering an alternative to detector pair-differencing, and thereby avoid systematics related to the differential properties of detector pairs\cite{Shimon:2007au}. It also mitigates the effects of $1/f$ noise from long-timescale atmospheric variations and reduces optical $I \rightarrow P$ leakage~\cite{Kusaka14,Takakura:2017ddx}. 

However, HWPs can also be a source of systematic effects like the HWPSS. Additionally, the HWPSS can cause non-linearity in the detector response, introducing another source of $I\rightarrow P$ leakage, which is not the dominant cause of the leakage. In this work, we have studied some systematic effects connected with the planned use of a CRHWP on the SO SAT. The aim of this study is to highlight our ability to characterize these effects so we can develop efficient mitigation strategies. We build from existing experimental efforts that have demonstrated the advantages of the use of HWPs as polarization modulators.

In this work, we used the Generalized Transfer Matrix method to 
calculate the Mueller matrix of a HWP stack at an arbitrary incidence angle,
and have analyzed multiple sources of the HWPSS.
Based on an illustrative model for the optical chain, we have estimated the amplitudes of the steady $2f$ and $4f$ components, and the $I\rightarrow P$ leakage coefficients.
We have shown that for an SO-like SAT system with a cold rotating HWP, the dominant sources of $I\rightarrow P$ leakage are the differential transmission of the alumina filter in front of the HWP, and modulation of unpolarized light through the HWP at large incidence angles. We also show that the $I \rightarrow P$ from  $0f \rightarrow 4f$ coupling through detector non-linearity, differential transmission through the window, and differential reflection off of the sky-side alumina filter are non-negligible. When combining the HWPSS with nonlinearity in simulations for the SAT optical setup, we observe a total leakage coefficient of $0.78\%$ for detectors at the edge of the focal plane, 
and $0.02\%$ for a detector at the center of the focal plane. We note that this is without any $I\rightarrow P$ removal so it represents an upper limit.

While the Generalized Transfer Matrix Method is widely used to study the systematic effects of HWPs, this is the first study to use it to both predict the HWPSS at such large angles of incidence for an achromatic HWP and to couple it with detector non-linearity-induced effects. We have also presented the calculation of Mueller matrix elements for meta-material silicon HWP (Appendix\,\ref{sec:metamaterial}). We also highlight the importance of considering slant incidence cases. The HWPSS estimation methods used here will inform the SO design and the development of dedicated mitigation strategies for HWP-related systematic effects. It will be also applicable to future experiments that consider employing CRHWPs.  
Future work will include the final SO HWP design and the propagation of the impact of the HWP $I\rightarrow P$ on maps, CMB spectra, and the estimation of cosmological parameters.


\acknowledgments 
This work was supported in part by a grant from the Simons Foundation (Award \#457687, B.K.). 
We would like to thank Tom Essinger-Hileman for many helpful discussions and for making his TMM code publicly available. The authors also acknowledge Julien Peloton for the development of the \texttt{s4cmb} pipeline and his instruction on how to use it.  MS acknowledges the financial support of the UnivEarthS Labex program at Sorbonne Paris Cit\'{e} (ANR-10-LABX-0023 and ANR-11-IDEX-0005-02). MG and JEG acknowledge support by the Vetenskapsr\aa det (Swedish Research Council) through contract No. 638-2013-8993 and the Oskar Klein Centre for Cosmoparticle Physics.


\appendix
\section{Meta-material HWPs}\label{sec:metamaterial}

\begin{figure}
 \begin{center}
 \includegraphics[width=0.5\textwidth]{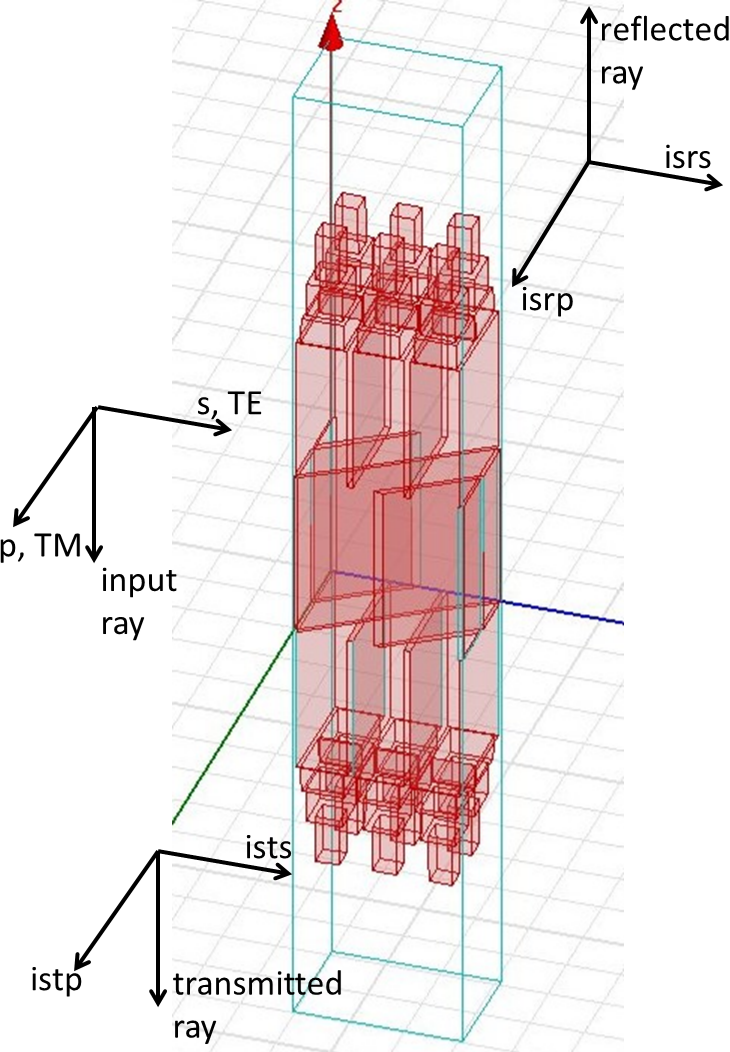}
 \end{center}
 \caption{Geometry of the HFSS simulation of the meta-material HWP. The case of an incoming $s$ polarization is shown. There is a similar trend for an incoming $p$ polarization. Geometry of the HFSS simulations from Coughlin 2018~\cite{Coughlin18}.}\label{fig:Si_geometry}
 \end{figure}

To model a meta-material HWP with a meta-material AR coating, we use HFSS simulations. We consider the  Advanced ACTPol (AdvACT) HWP design optimized for 90~GHz and 150~GHz~\cite{Henderson16,Coughlin18}. By applying suitable electromagnetic boundary conditions, it is possible to simulate one unit of meta-material. The use of Floquet Ports~\cite{floquet} enables this unit to exhibit the behavior representative of the entire device. We combine the typical outputs of the HFSS simulations, i.e. the real and imaginary parts of the electric fields transmitted and reflected along two reference orthogonal directions $s$ and $p$, $\Re_{ijlm}$ and $\Im_{ijlm}$ with $i,j,l,m=1,2$ the index of the HFSS Floquet ports\cite{floquet}. Suitable values of these indices give electric fields along the $s$ and $p$ directions to build the adimensional intensities along the two HWP axes. For example, if $\Re_{1121}$ ($\Re_{1221}$) is the real part of the amplitude of the electric field transmitted on the $s$ ($s$) axis from an input polarization along the $s$ ($p$) axis, the amplitude of the electric field transmitted on the $s$ axis from an incoming polarization along the $s$ and $p$ axis are, respectively (see also Fig.\,\ref{fig:Si_geometry}): 
\begin{eqnarray}\label{e2}
\nonumber ists & = & \sqrt{\Re_{1121}^2+\Im_{1121}^2}, \\ 
          ipts & = & \sqrt{\Re_{1221}^2+\Im_{1221}^2}.
\end{eqnarray}
From these quantities, the total adimensional intensity $Ts$ along the $s$ axis can be written as: $Ts= ists^2 + ipts^2$. Similarly,  the total and the differential intensities in transmission and in reflection (i.e. the top left part of the Mueller matrix) can be built from the HFSS output electric field amplitudes. Differential phase delays are estimated as:
\begin{eqnarray}\label{e3}
\Gamma=\Gamma_s+\Gamma_p=\arctan{(\frac{istp}{ists}+\frac{\pi}{2})} + \arctan{(\frac{iptp}{ipts})}.
\end{eqnarray}
$\Gamma_s (\Gamma_p)$ is the phase delay along the $s (p)$ axis, their combined action provides the overall HWP phase delay. Eq.~\ref{e3} works for the phase delay both in transmission and in reflection.

\begin{figure}
 \begin{center}
 \includegraphics[width=0.7\textwidth]{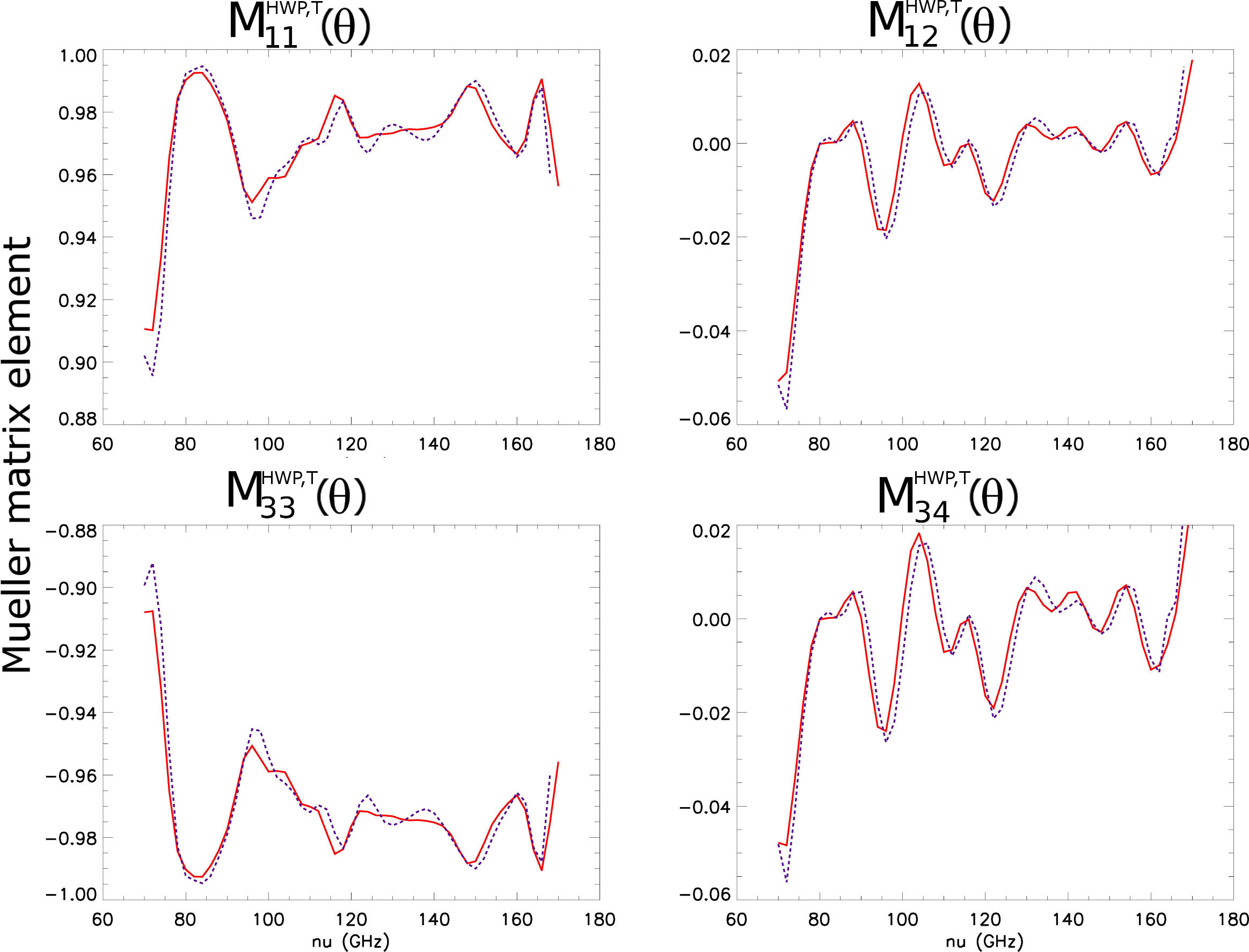}
 \end{center}
 \caption{Mueller matrix non-zero elements of the AdvACT meta-material silicon AHWP in transmission as a function of the frequency $\nu$ of the incident wave, for $\theta=0^{\circ}$ (red solid line) and $\theta=20^{\circ}$ (blue dashed line).}\label{fig:Si_M^T_ij_0-20deg}
 \end{figure}

\begin{figure}
 \begin{center}
 \includegraphics[width=0.7\textwidth]{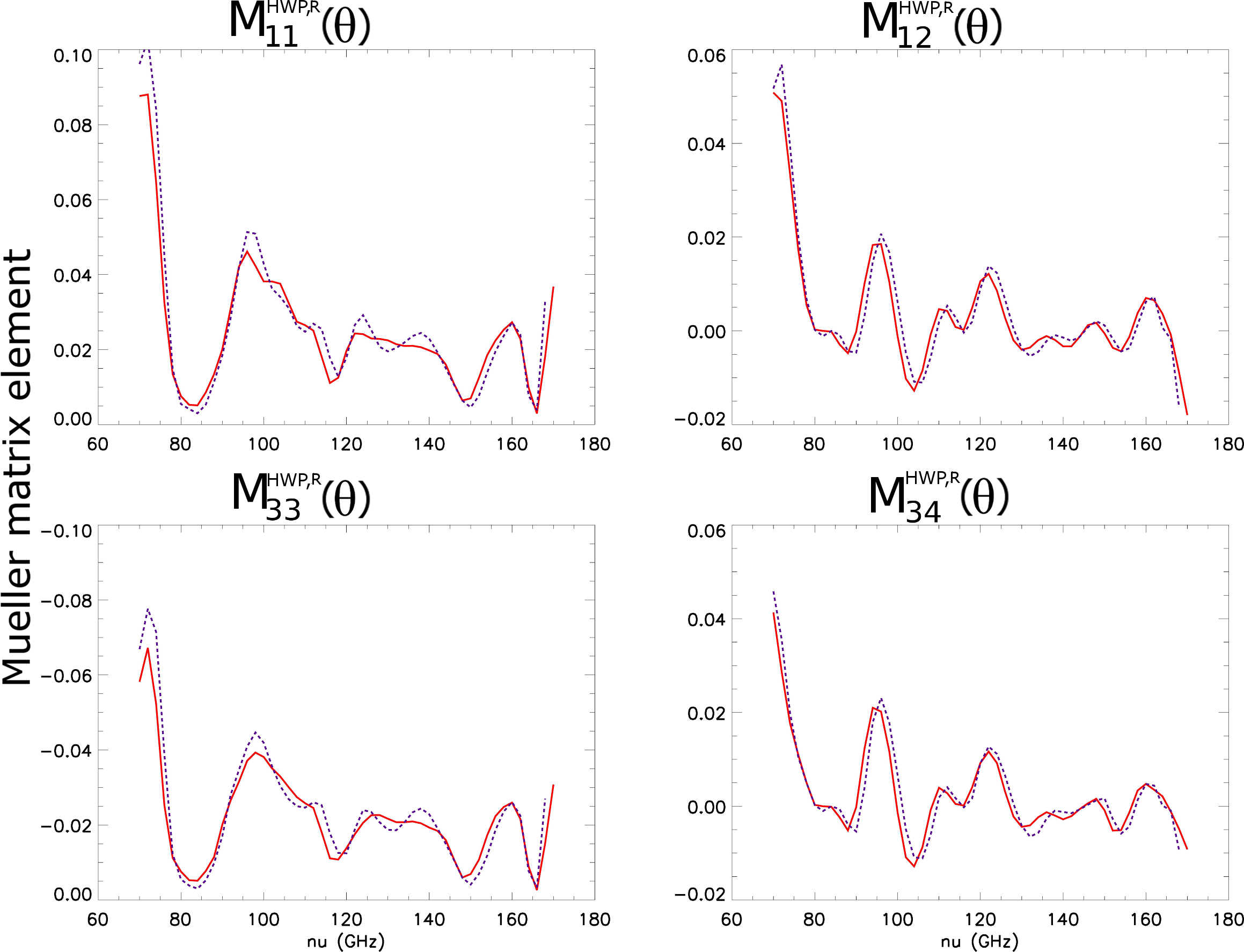}
 \end{center}
 \caption{Mueller matrix non-zero elements of the AdvACT meta-material silicon AHWP in reflection as a function of the frequency $\nu$ of the incident wave, for $\theta=0^{\circ}$ (red solid line) and $\theta=20^{\circ}$ (blue dashed line).}\label{fig:Si_M^R_ij_0-20deg}
 \end{figure}

Intensities (Eq.\,\ref{e2}) and phase delays (Eq.\,\ref{e3}) along the two HWP axes are used to build the meta-material HWP Mueller matrices in transmission (Fig.\,\ref{fig:Si_M^T_ij_0-20deg}) and in reflection (Fig.\,\ref{fig:Si_M^R_ij_0-20deg}) for incident angles $\theta=0^{\circ}$ and $\theta=20^{\circ}$  following Eq.~1 in Zhang et al.~\cite{Zhang11} and Savini et al.~\cite{Savini09}. Similarly to the sapphire HWP (Figs.\,\ref{fig:Mij_diff_one}, \ref{fig:Mij_diff_three90} and \ref{fig:Mij_diff_three150}), the difference between $\theta=0^{\circ}$ and $\theta=20^{\circ}$ is very small.

While a rotating sapphire HWP is traditionally derived by multiplying the Mueller matrix of the static HWP by the rotation matrices, the meta-material HWP requires HFSS simulations of the HWP rotated at different angles $\chi$.

 \begin{figure}
 \begin{center}
 \includegraphics[width=0.7\textwidth]{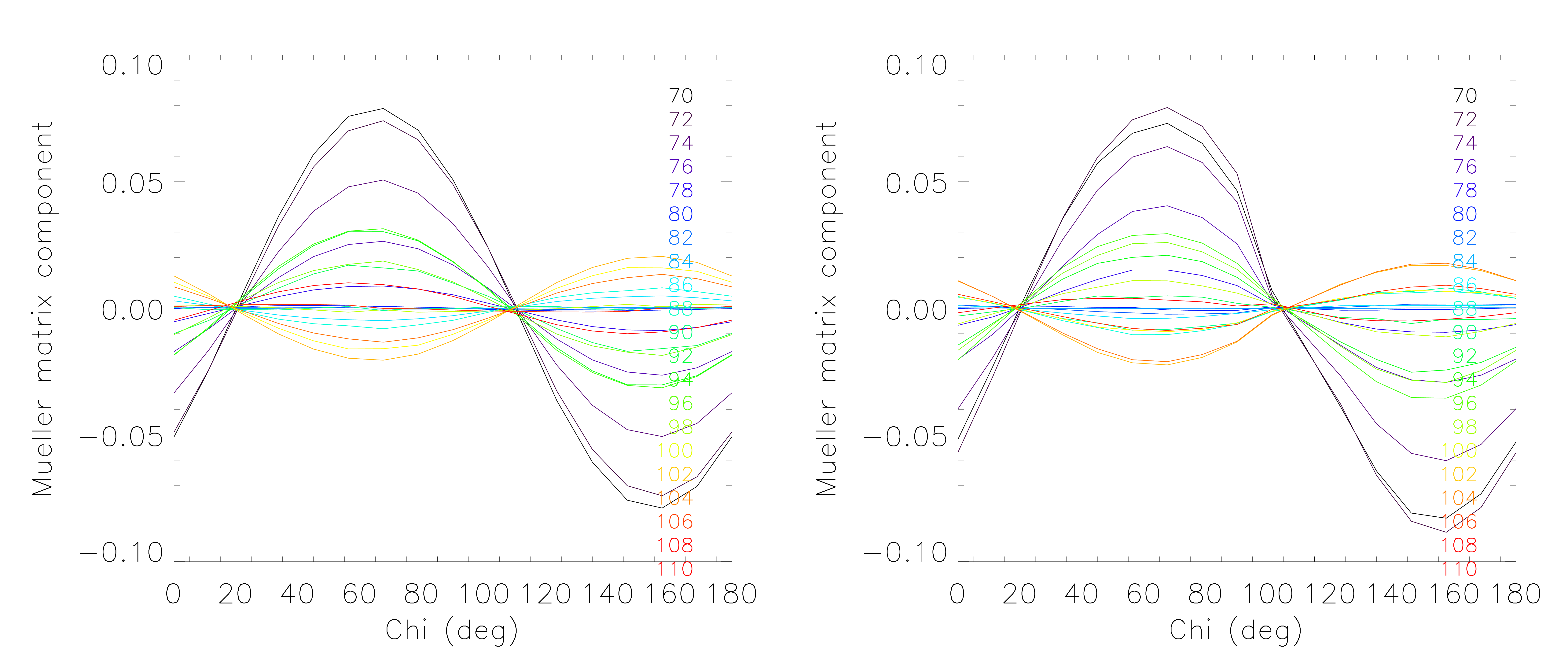}
 \end{center}
 \caption{$M_{12}$ Mueller matrix elements of the AdvACT meta-material Silicon AHWP
 as a function of the rotating angle $\chi$ for $\theta=0^{\circ}$ (left) and 20$^{\circ}$ (right). The $M_{12}$ trend is derived using HFSS simulations rather than the traditional multiplication of rotation matrices.}\label{fig:Si_Mij_theta_0-20deg}
 \end{figure}

\nocite{*} 
\bibliography{report} 
\bibliographystyle{spiebib} 

\end{document}